\documentclass[twocolumn,showpacs,preprintnumbers,amsmath,amssymb]{revtex4}
\usepackage{graphics}
\usepackage{graphicx}
\begin{document}

\title{Quantum Correlation in One-dimensional Extended Quantum Compass Model}
\author{Wen-Long You}
\altaffiliation{Email: wlyou@suda.edu.cn} \affiliation{School of
Physical Science and Technology, Soochow University, Suzhou, Jiangsu
215006, People's Republic of China}
\date{\today}
\begin{abstract}
We study the correlations in the one-dimensional extended quantum
compass model in a transverse magnetic field. By exactly solving the
Hamiltonian, we find that the quantum correlation of the ground
state of one-dimensional quantum compass model is vanishing. We show
that quantum discord can not only locate the quantum critical
points, but also discern the orders of phase transitions.
Furthermore, entanglement quantified by concurrence is also
compared.
\end{abstract}


\maketitle
\section{Introduction}
\label{Introduction} Recently, there has been a revived interest in
the study of correlations in quantum systems. In the framework of
quantum-classical dichotomy, the total correlations can be separated
into a purely quantum part and a classical counterpart, and
both quantum correlation and classical correlation are able to be quantified
respectively. A quantitative understanding of the different types of
correlations might aid the application of quantum manipulation
\cite{Henderson}. Therefore, distinguishing classical and quantum
correlations in quantum systems is of both fundamental and practical
importance. Paradoxically, entanglement was considered to be the most suitable manifestation of the quantum correlation and
the main resource that speedup quantum computers over their classical counterparts \cite{Amico,Horodecki}. The studies of quantum correlations in exactly solvable models
has a long tradition. For instance, the open-system dynamics of correlations in the presence of effect of environment is explored  based on exact description \cite{PhysRevA.77.042316,EPJD.57.439,PhysRevA.84.032120}.

Let us consider a bipartite system described by the
density operator $\rho_{AB}$ shared by parts $A$ and $B$, and if it
can be written in the separable form $ \rho_{AB}=\sum_{j} p_j
\rho_{A}^{(j)} \otimes \rho_{B}^{(j)}$, such mixed bipartite state
is termed disentangled. Though all the correlations therein are local, some separable quantum states still contain intrinsically
quantum correlations \cite{White}. In this sense, it seems entanglement is not
always needed for quantum speed-ups \cite{Datta}. Quantum discord
(QD) was thus introduced to quantify non-classical correlations
beyond entanglement paradigm in quantum states
\cite{PhysRevLett.88.017901}, and it has received an astonishingly
amount of interest both theoretically and experimentally
\cite{PhysRevA.81.062118,Nat.Commun.1.7,PhysRevA.81.042105,Auccaise,arXiv1112.6238}.

In the field of quantum information, for a bipartite system
$\rho_{AB}$, the total mutual information is the relative entropy
between $\rho_{AB}$ and $\rho_{A} \otimes \rho_{B}$, which
corresponds to the minimal rate of randomness that is required to
completely erase all the correlations in $\rho_{AB}$,
\begin{eqnarray}
I(\rho_{AB}) =S(\rho_{A}) + S(\rho_{B})-S(\rho_{AB}),
\end{eqnarray}
with von Neumann entropy $S(\rho) = -\textrm{Tr}{\rho} \log(\rho)$.
The quantum conditional entropy over a set of Von Neumann
measurement $B_k$ is defined by
\begin{eqnarray}
S(\rho_{AB}|{B_k}) :=\sum_{k} p_k S(\rho_k),
\end{eqnarray}
where the measurement-based conditional density operator $\rho_k$
associated with the measurement result $k$ is
\begin{eqnarray}
\rho_{k}=\frac{1}{p_{k}}(I_A \otimes B_k)\rho_{AB} (I_A\otimes B_k),
\end{eqnarray}
in which $I_A$ is the identity operator of the subsystem $A$ and
$p_{k}$= tr$[(I_A\otimes B_k)\rho_{AB} (I_A\otimes B_k)]$.
Consequently, the associated quantum mutual information is given by
\begin{eqnarray}
I(\rho_{AB}|{B_k})= S(\rho_A)-S(\rho_{AB}|{B_k}).
\end{eqnarray}
The classical mutual correlation is defined as the superior of
$I(\rho_{AB}|{B_k})$ over all possible sets of one-dimensional
positive-operator-valued measure (POVM) $B_k$,
\begin{eqnarray}
C(\rho_{AB})=\sup_{\{B_k\}} I(\rho_{AB}|{B_k}).
\label{classicalmutualcorrelation}
\end{eqnarray}
The QD is then given by the difference of mutual information
$I(\rho_{AB})$ and the classical correlation $C(\rho_{AB})$,
\begin{eqnarray}
D(\rho_{AB})=I(\rho_{AB})-C(\rho_{AB}).
\end{eqnarray}
Since the minimization taken over POVMs is a notorious problem, so
far only a few analytical results are obtained, including the
Bell-diagonal states \cite{PhysRevA.77.042303,Lang}, rank-2 states
\cite{Cen}, and Gaussian states \cite{PhysRevLett.105.030501,Giorda}.
 Numerical efforts should be desired for general states. The QD can be shown to be asymmetric and nonnegative
\cite{RevModPhys.75.715}, and it is invariant
under local unitary transformations. The QD will vanish if and only
if the state is classical, and it is
implied that classical-only correlated quantum states are extremely
rare \cite{PhysRevA.81.052318}.

The QD not only can discern classical and quantum correlations, but
also can be used to establish relation to quantum phase transitions
(QPTs) of many-body systems. Information on the locations and the orders
of the QPTs can be obtained by consideration of the derivatives of
the bipartite QD with respect to the coupling parameters
\cite{PhysRevB.78.224413}, and has also been generalized to
multipartite state \cite{PhysRevLett.104.080501,Parashar}. As we
know, a QPT identifies any point of nonanalyticity in the
ground-state (GS) energy of an infinite lattice system. The patterns
in correlations of a many-body system suddenly change across the quantum
critical point (QCP), and induce the non-analytic behavior of
ground state $\vert \Psi_0\rangle$. The reduced density matrix (RDM)
$\rho_{ij}= \textrm{Tr}_{ij} \vert \Psi_0\rangle\langle \Psi_0
\vert$ is obtained by taking a partial trace over all degrees of
freedom except particles $i$ and $j$. The information of QPT is
encoded in the nonanalyticity of the matrix elements. With the QD being calculated from the RDM,
one deduces that a discontinuity in the QD implies a first-order
QPT, and a singularity in the derivative of the QD implies a second-order QPT \cite{PhysRevLett.93.250404}. Hence, the quantum correlation can
serve as a hallmark for the QPT in the many-body system. We would like
to stress that the RDM comprises more accessible information than the
information of QPT only.

In this respect, we will take advantage of QD to study
one-dimensional (1D) extended quantum compass model (EQCM) in the
transverse magnetic field. The main reason for focusing on this
model is that it not only allow us to study second-order phase
transitions, but also allow the investigation of first-order
transitions. The rest of the paper is organized as follows. In Sec.
\ref{modelsolution}, we introduce the 1D EQCM in the transverse
magnetic field, and exploit the exact solutions. In Sec.
\ref{quantumdiscordofmodel} we calculate two-qubit QD of 1D EQCM. We
show that QD can not only diagnose various phase transitions, but
also identify the character of the QPTs. Consequently, the zero
temperature phase diagram of the model is mapped out. In Sec.
\ref{Entanglement}, we recheck the entanglement in the 1D EQCM in
terms of concurrence. Sec. \ref{Conclusion} finally contains the comparison between
the concurrence and the QD and a short summary.

\section{Quantum phase transition in one-dimensional extended quantum compass
model in a transverse field} \label{modelsolution} The Hamiltonian
of 1D EQCM in an external transverse magnetic field is given by
\cite{Brzezicki1,Brzezicki2}
\begin{eqnarray}
H&=&\sum_{i=1}^{N^\prime}[J_1 \sigma_{2i-1}^x \sigma_{2i}^x + J_2
\sigma_{2i-1}^y \sigma_{2i}^y + L_1 \sigma_{2i}^x \sigma_{2i+1}^x \nonumber \\
&+& L_2 \sigma_{2i}^y \sigma_{2i+1}^y + \frac{h}{2}
({\sigma}_{2i-1}^z + {\sigma}_{2i}^z)], \label{Hamiltonian3}
\end{eqnarray}
where $\sigma_i^{a}$ $(a=x,y,z) $ is the Pauli operator at site
$\textit{i}$, $J_1$ and $J_2$ ( $L_1$ and $L_2$) are the
strength of the nearest-neighbor interaction on the odd (even) bond,
and $h$ characterizes the intensity of the external field applied in
the $z$ direction. $N=2N^\prime$ is the number of the sites. We
assume cyclic boundary conditions, i.e., the ($N$ + 1)th site is
identified with the first site. The Hamiltonian describes a
structure of two spins inside a unit cell. The orbital-orbital
interactions depend strongly on the bond between two neighboring
lattice sites. The Hamiltonian (\ref{Hamiltonian3}) encompasses two
other well-known spin models: it turns into transverse Ising chain
for $J_1=L_1, J_2=L_2=0$ and the XY chain in a transverse field for
$J_1=L_1, J_2=L_2$.

The Hamiltonian (\ref{Hamiltonian3}) can be exactly diagonalized by following the standard procedures. The Jordan-Wigner transformation
maps explicitly between spin operators and spinless fermion
operators by \cite{Jordan-Zphys-1928,Sachdev}
\begin{eqnarray}
\sigma _{j}^{+}& =&\exp \left[ i \pi \sum_{i=1}^{j-1}c_{i}^{\dagger }c_{i}%
\right] c_{j}=\prod_{i=1}^{j-1}\sigma _{i}^{z}c_{j},  \notag \\
\sigma _{j}^{-}& =&\exp \left[ -i\pi \sum_{i=1}^{j-1}c_{i}^{\dagger }c_{i}%
\right] c_{j}^{\dagger }=\prod_{i=1}^{j-1}\sigma
_{i}^{z}c_{j}^{\dagger },
\notag \\
\sigma _{j}^{z}& =&1-2c_{j}^{\dagger }c_{j}.
\end{eqnarray}%
Next discrete Fourier transformation for plural spin sites is
introduced by
\begin{eqnarray}
c_{2j-1}=\frac{1}{\sqrt{N'}}\sum_{k}e^{-ik j}a_{k},\text{ \ \ }c_{2j}=%
\frac{1}{\sqrt{N'}}\sum_{k}e^{-ik j}b_{k},
\end{eqnarray}
with the discrete momentums as
\begin{eqnarray}
k=\frac{n\pi}{ N^\prime  }, \quad n= -(N^\prime-1), -(N^\prime-3),
\ldots, N^\prime -1.
\end{eqnarray}
Finally, the diagonalized form is achieved by a four-dimensional
Bogoliubov transformation with two kind of quasiparticles
\cite{Brzezicki2,arXiv1101.3673},
\begin{eqnarray}
H=\sum_{k} \left[E_k^o (\gamma_k^{o\dagger}\gamma_k^{o}
-\frac{1}{2}) +  E_k^a (\gamma_k^{a\dagger}\gamma_k^{a}
-\frac{1}{2}) \right],
\end{eqnarray}
where optical spectra $E_k^o=\sqrt{\varsigma+\sqrt{\tau}}$ and
acoustic spectra $E_k^a=\sqrt{\varsigma-\sqrt{\tau}}$. Here
$\varsigma=\vert \alpha_k \vert^2 + \vert \beta_k \vert^2$,
$\tau=(\alpha_k^* \beta_k + \alpha_k \beta_k^*)^2 + 4 \vert \alpha_k
\vert^2 h^2$, $\alpha_k= (J_1+J_2) + (L_1 + L_2) e^{ik}$, and
$\beta_k= (J_1 -J_2)- (L_1 -L_2) e^{ik}$. The ground state $E_0$ is
obtained,
\begin{eqnarray}
E_0 = -\frac{1}{2} \sum_{k} \left(E_k^o + E_k^a\right).
\end{eqnarray}
It is easy to find that the energy gap of acoustic branch will close
when $h=2\sqrt{(J_1\pm L_2)(J_2 \pm  L_1)}$, and the
nonanalyticities of the ground state determine QCPs. In the absence
of the transverse magnetic field, i.e., $h$ = 0, the critical lines
correspond to $J_1= \pm L_2$ and $J_2=\pm L_1$, respectively, which
confirms the conclusion in Ref. \cite{PhysRevB.80.174417}.

\section{Two-Qubit quantum discord of extended quantum compass model}
\label{quantumdiscordofmodel} The QD is explored from the two-qubit
RDM. In the representation spanned by the two-qubit product
states $\{ \vert 0 \rangle_A \otimes \vert 0 \rangle_B$, $ \vert 0
\rangle_A \otimes \vert 1 \rangle_B$, $ \vert 1 \rangle_A  \otimes
\vert 0 \rangle_B$, $\vert 1 \rangle_A \otimes \vert 1 \rangle_B
\}$, where $\vert 0 \rangle$ ($\vert 1 \rangle$) denotes spin up (down) state, the two-site density matrix can be expressed as,
\begin{equation}
\rho_{ij}=\frac{1}{4}\sum_{a,a'=0}^3\langle
\sigma_i^{a}\sigma_j^{a'}\rangle
   \sigma_i^{a} \sigma_j^{a'},
\end{equation}
where $\sigma_i^{a}$ are Pauli matrices $\sigma_i^{x}$,
$\sigma_i^{y}$ and $\sigma_i^{z}$ for $a$ = 1 to 3, and 2 by 2 unit
matrix for $a$=0. The Hamiltonian has $Z_2$ symmetry, namely, the
invariance under parity transformation $P=\otimes_{i} \sigma_{i}^z$,
and then correlation functions such as $\langle
\sigma_i^{a}\sigma_j^{b}\rangle$ ($a=x,y$ and $b=0,z$) simultaneously
vanish. Also, $\langle \sigma_i^{x}\sigma_j^{y}\rangle$  ($\langle \sigma_i^{y}\sigma_j^{x}\rangle$) is zero due
to the imaginary character of $\sigma_j^{y}$ ($\sigma_i^{y}$). Therefore, the
two-qubit density matrix reduces to a X-state ,
\begin{equation}
\rho_{ij}=\left(
\begin{array}{cccc}
u^{+} & 0 & 0 & z^{-} \\
0 & w_{1} & z^{+} & 0 \\
0 & z^{+} & w_{2} & 0 \\
z^{-} & 0 & 0 & u^{-}%
\end{array}%
\right),  \label{eq:2DXXZ_RDM}
\end{equation}%
with
\begin{eqnarray}
&&u^{\pm }=\frac{1}{4}(1\pm  2\langle {\sigma_{i}^{z}}\rangle
+\langle {\sigma _{i}^{z}\sigma _{j}^{z}}\rangle ),
\label{upm} \\
&&z^{\pm }=\frac{1}{4}(\langle \sigma _{i}^{x}\sigma _{j
}^{x}\rangle \pm \langle \sigma _{i}^{y}\sigma _{j
}^{y}\rangle ),  \label{zpm} \\
&&\omega _{1}=\omega _{2}=\frac{1}{4}(1-\langle \sigma _{i
}^{z}\sigma _{j}^{z}\rangle ).  \label{omega1}
\end{eqnarray}%
The density matrix of single qubit is easily obtained by a partial trace over
one of the two qubits,
\begin{equation}
\rho_{i}=\left(
\begin{array}{c c}
\frac{1}{2}(1+\langle {\sigma_{i}^{z}}\rangle) & 0  \\
0 & \frac{1}{2} (1- \langle {\sigma_{i}^{z}}\rangle)
\end{array}%
\right) .  \label{eq:2DXXZ_RDM}
\end{equation}%
Thus the total correlation is quantified by the quantum mutual
information as
\begin{equation}
I(\rho_{ij})=S(\rho_{i})+S(\rho_{j})-S(\rho_{ij}), \label{qminfo}
\end{equation}
with $S(\rho_{i})$=$S(\rho_{j})$= $-\sum_{m=0}^1$ $\{ [1+(-1)^m \langle
\sigma_{i}^z \rangle ]/2\}$$\log_2\{ [1+(-1)^m \langle \sigma_{i}^z
\rangle ]/2\}$ and $ S(\rho_{ij})$=$-\sum_{m=0}^1$ $\xi_m$ $\log_2 \xi_m$
$-\sum_{n=0}^1$ $\xi_n$ $\log_2 \xi_n$,
where $\xi_m$ = $[1+ \langle
\sigma_{i}^z \sigma_{j}^z \rangle$ + $(-1)^m
\sqrt{(\langle\sigma_{i}^x \sigma_{j}^x\rangle-\langle\sigma_{i}^y
\sigma_{j}^y\rangle)^2 + 4 \langle \sigma_{i}^z \rangle^2 }]/4$ and
$\xi_n = [1 - \langle \sigma_{i}^z \sigma_{j}^z \rangle +
(-1)^n(\langle\sigma_{i}^x \sigma_{j}^x\rangle+\langle\sigma_{i}^y
\sigma_{j}^y\rangle) ]/4$. Since part $B$ contains a single qubit,
we can compute the classical correlation by extremizing Eq.
(\ref{classicalmutualcorrelation}) over a complete set of orthogonal
projectors $\{B_k=\vert \Theta_\kappa \rangle \langle \Theta_\kappa
\vert$, $\kappa= \parallel, \perp \}$, where $\Theta_\parallel
\equiv \cos (\theta/2) \vert 0 \rangle_B+ e^{i \varphi} \sin
(\theta/2) \vert 1 \rangle_B$ and $\Theta_\perp \equiv e^{-i
\varphi} \sin (\theta/2) \vert 0 \rangle_B -   \cos(\theta/2) \vert
1 \rangle_B$ with $0 \le \theta \le \pi$ and $0 \le \varphi < 2
\pi$. Extensive numerical analysis implies that the extremization is
mostly achieved at $\theta=\pi/2$, $\phi=0$
\cite{PhysRevA.80.022108,XGWang}, which is also confirmed in our
numerical optimization. Accordingly, the classical correlation is
expressed as
\begin{eqnarray}
C(\rho_{ij}) = {H}_{\textrm{bin}}(p_1)- {H}_{\textrm{bin}}(p_2),
\label{Classicalcorrelation1}
\end{eqnarray}
where ${H}_{\textrm{bin}}(p)=-p\log(p)-(1-p)\log(1-p)$ is the binary
entropy, $p_1 = (1+  \langle \sigma_{i}^z \rangle)/2$ and $
p_2 = \left(1+  \sqrt{[\max (\vert \langle \sigma_{i}^x
\sigma_{j}^x \rangle \vert, \vert\langle \sigma_{i}^y \sigma_{j}^y
\rangle \vert)]^2 +   \langle \sigma_{i}^z \rangle^2 }\right)/2$. Thus the
quantum correlation is simply given by
\begin{eqnarray}
D(\rho_{ij})=I(\rho_{ij})-C(\rho_{ij}).\label{qcorr}
\end{eqnarray}
The Eq.(\ref{Classicalcorrelation1}) has been verified in Werner
state \cite{PhysRevA.77.042303}, 1D Ising model, XY model
\cite{PhysRevA.82.012106}, and XXZ model
\cite{PhysRevA.80.022108,PhysRevB.78.224413}, even for finite
temperature \cite{PhysRevA.82.012106,PhysRevLett.105.095702,PhysRevA.83.062334}.

\begin{figure}[t]
\begin{center}
\includegraphics[width=8cm]{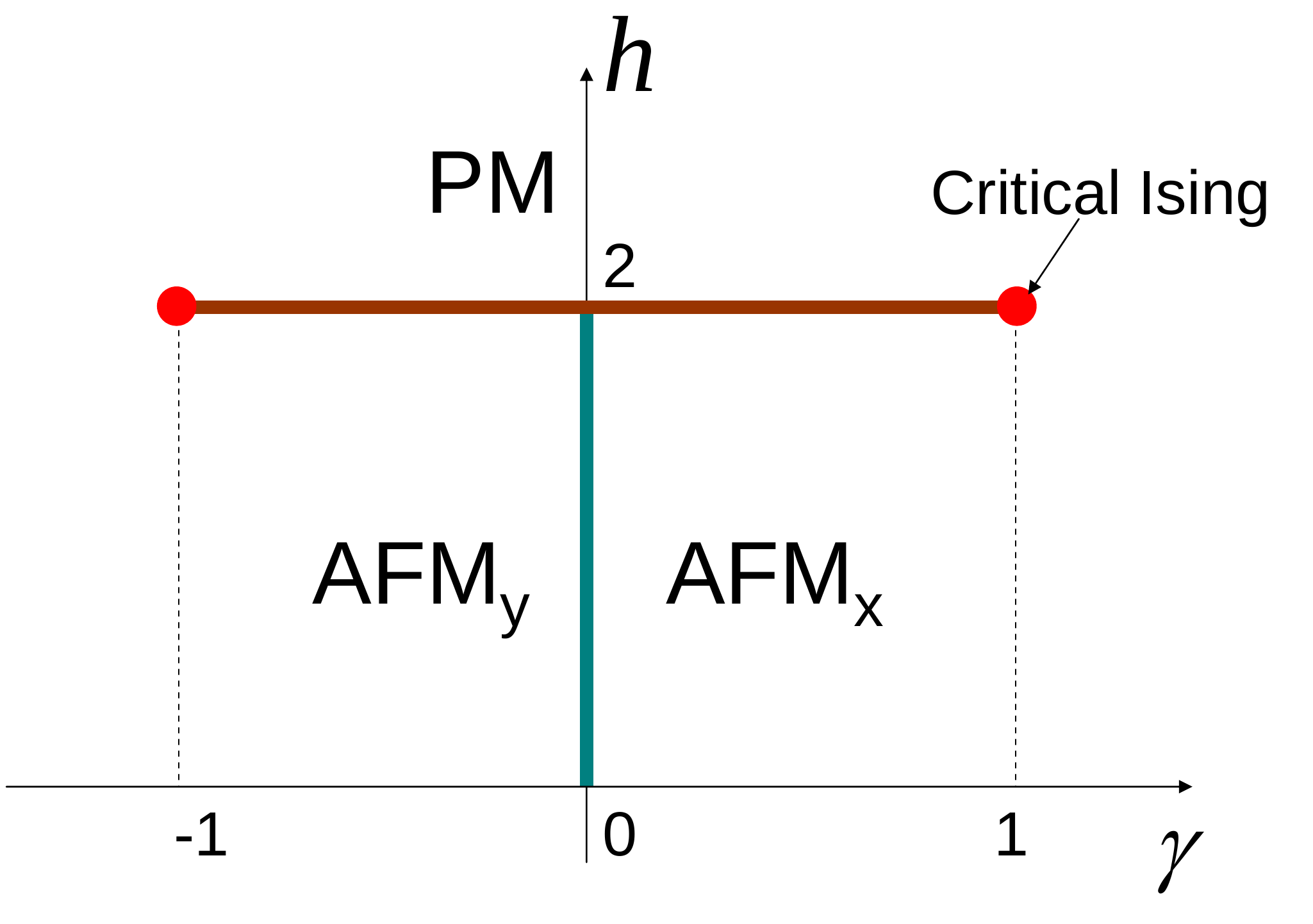} %
\includegraphics[width=8cm]{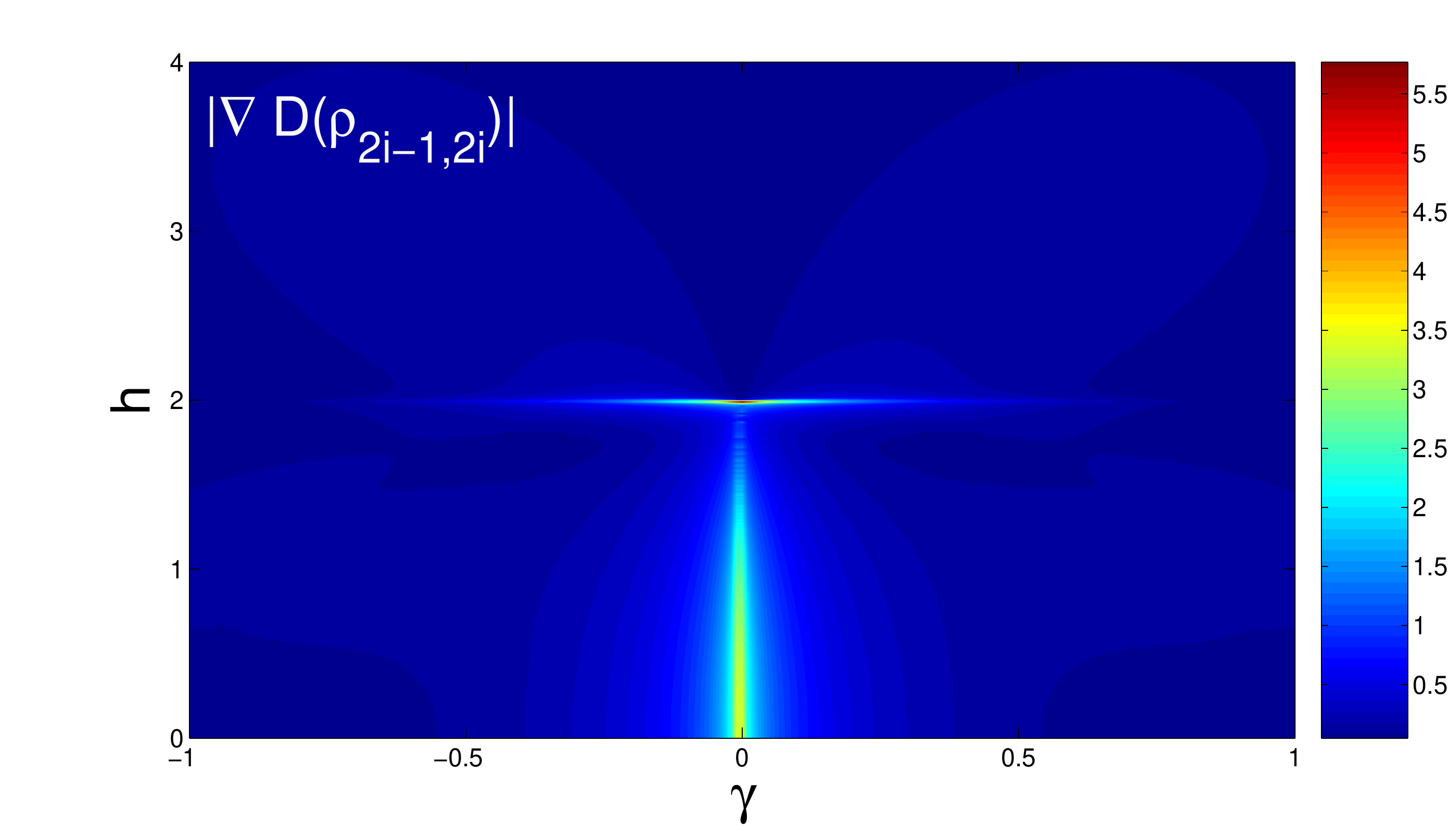}
\caption{(Color online) Top: phase diagram for the anisotropic XY
spin-1/2 chain in a transverse field at zero temperature. The heavy
lines represent second order phase transitions. The horizontal line
will be referred to as the anisotropic transition and the vertical
line as the Ising transition. PM denotes a paramagnetic phase and
AFM$_x$ (AFM$_y$) denotes antiferromagnetic phase along the $x$
($y$) direction. The XX model obtained by setting $\gamma$ = 0
displays a critical line for magnetic field $h \in$ [0,2]. The Ising
model obtained for $\gamma$ = 1 exhibits a critical point at $h$ =
2. Bottom: the first derivative of the QD between nearest neighbors
in the $\gamma-h$ plane.} \label{PhasediagramofXY}
\end{center}
\end{figure}

First, we consider the QPT in 1D XY model, i.e., $J_1=L_1$,
$J_2=L_2$. $\gamma=(J_1-J_2)/(J_1+J_2)$ denotes the anisotropy of
the coupling. The phase diagram is shown in Fig. \ref{PhasediagramofXY}, which describes two distinct QPTs. The
transverse magnetic field $h$ drives a second-order transition from
an antiferromagnetic ordered phase to a paramagnetic
quantum-disordered phase, and the anisotropy critical line
$\gamma=0$ is the boundary between a N\'{e}el state along the $x$
direction and a N\'{e}el state along the $y$ direction
\cite{Jacobson,Bunder}. The first derivative of the QD for the
nearest-neighboring odd bond is displayed in Fig. \ref{PhasediagramofXY}.
 It is evident there are rapid changes of QD in the critical regions. The changing rate of the QD is most pronounced for
critical line $\gamma=0$. The diverging first derivative of QD
suggests both transitions belong to second-order phase transitions.
Finite size scaling shows that the peak will not increase with
respect to lattice size.

Next, we consider the QPT in 1D compass model, i.e., $J_1=L_2=0$.
The ground state of finite-size system is $2^{N^\prime-1}$ fold degenerate \cite{You}. In the absence
of magnetic field $h=0$, we note that there is only classical
correlation created between nearest neighbors. In this case,
$E_k^o=2 \sqrt{J_2^2+L_1^2+2J_2L_1\cos k}$, $E_k^a=0$. The only
correlation functions surviving are $\mathcal{C}_{2i-1,2i}^{y}
\equiv \langle \sigma_{2i-1}^{y} \sigma_{2i}^{y} \rangle$ and
$\mathcal{C}_{2i,2i+1}^x \equiv \langle \sigma_{2i}^{x}
\sigma_{2i+1}^{x} \rangle$. As a consequence, the density matrix
becomes densities diagonal in the orthogonal product bases. In other
words, they are equivalent up to local unitary operations to
Bell-diagonal states \cite{Ciliberti},
\begin{eqnarray}
\rho_{2i-1,2i}&=&\frac{1}{4} \left(
\begin{array}{llll}
1 & 0 & 0 & -\mathcal{C}_{2i-1,2i}^{y} \\
0 & 1 & \mathcal{C}_{2i-1,2i}^{y} & 0 \\
0 & \mathcal{C}_{2i-1,2i}^{y} & 1 & 0 \\
-\mathcal{C}_{2i-1,2i}^{y} & 0 & 0 & 1%
\end{array}%
\right)\nonumber\\
&=&\sum_{j=1}^4 \lambda_j \vert \Psi_j \rangle \langle\Psi_j\vert,
\label{oddbondJ10L20}
\end{eqnarray}%
where $\vert \Psi_i\rangle$ are four Bell states with $\vert \Psi_1
\rangle =  (\vert 00 \rangle + \vert 11 \rangle)/\sqrt{2}$,
$\vert \Psi_2 \rangle = (\vert 01 \rangle + \vert
10 \rangle)/\sqrt{2}$, $\vert \Psi_3 \rangle = (\vert 01
\rangle - \vert 10 \rangle)/\sqrt{2}$, $\vert \Psi_4 \rangle =
(\vert 00 \rangle - \vert 11 \rangle)/\sqrt{2}$, and $\lambda_1=\lambda_3 = (1-\mathcal{C}_{2i-1,2i}^{y})/8$,
$\lambda_2=\lambda_4 = (1+\mathcal{C}_{2i-1,2i}^{y})/8$. We can find that QD vanishes in case of maximum $\lambda_i$ is less
than 0.5 \cite{PhysRevLett.104.080501,Huang}, that is, $D(\rho_{2i-1,2i})$=0.
Similarly, the quantum correlation between two qubits on even bonds
is also equal to zero, i.e., $D(\rho_{2i,2i+1})$=0. In a sense, 1D compass model behaves
as classical system. The null QD is induced by the macroscopic
degeneracy of the ground state dut to the peculiar symmetry of the
Hamiltonian \cite{Brzezicki}. The similar strategy applies for the case $J_2=0$, $L_2=0$, $h=0$. As a consequence, the energy spectra are greatly simplified as $E_k^o=2 J_1$ and
$E_k^a=2 L_1$. Then, the correlation functions are found to be $\langle \sigma_{2i-1}^y \sigma_{2i}^y \rangle$=$\langle \sigma_{2i}^y \sigma_{2i+1}^y
 \rangle=0$ and $\langle \sigma_{2i-1}^x \sigma_{2i}^x \rangle$= $\langle \sigma_{2i}^x
 \sigma_{2i+1}^x \rangle=-1$.  In such case, $D(\rho_{i,i+1})=0$. We have verified
that if a bipartite quantum state is a product state, i.e., $
\rho_{AB}= \rho_{A} \otimes \rho_{B}$, the state has no quantum
correlations, not vice versa.

It has been shown that zero QD between
a quantum system and its environment is necessary and sufficient for
describing the evolution of the system through a completely positive
map \cite{Shabani,Rodriguez}. In addition, a quantum state can be
locally broadcast, i.e., of locally
sharing preestablished correlations,  if and only if it has zero QD
\cite{Barnum,PhysRevLett.100.090502}. These purely classical states
are rather unsteady, and a generic arbitrarily
small perturbation will make them become nonclassical \cite{PhysRevA.81.052318}.

\begin{figure}[t]
\begin{center}
\includegraphics[width=8cm]{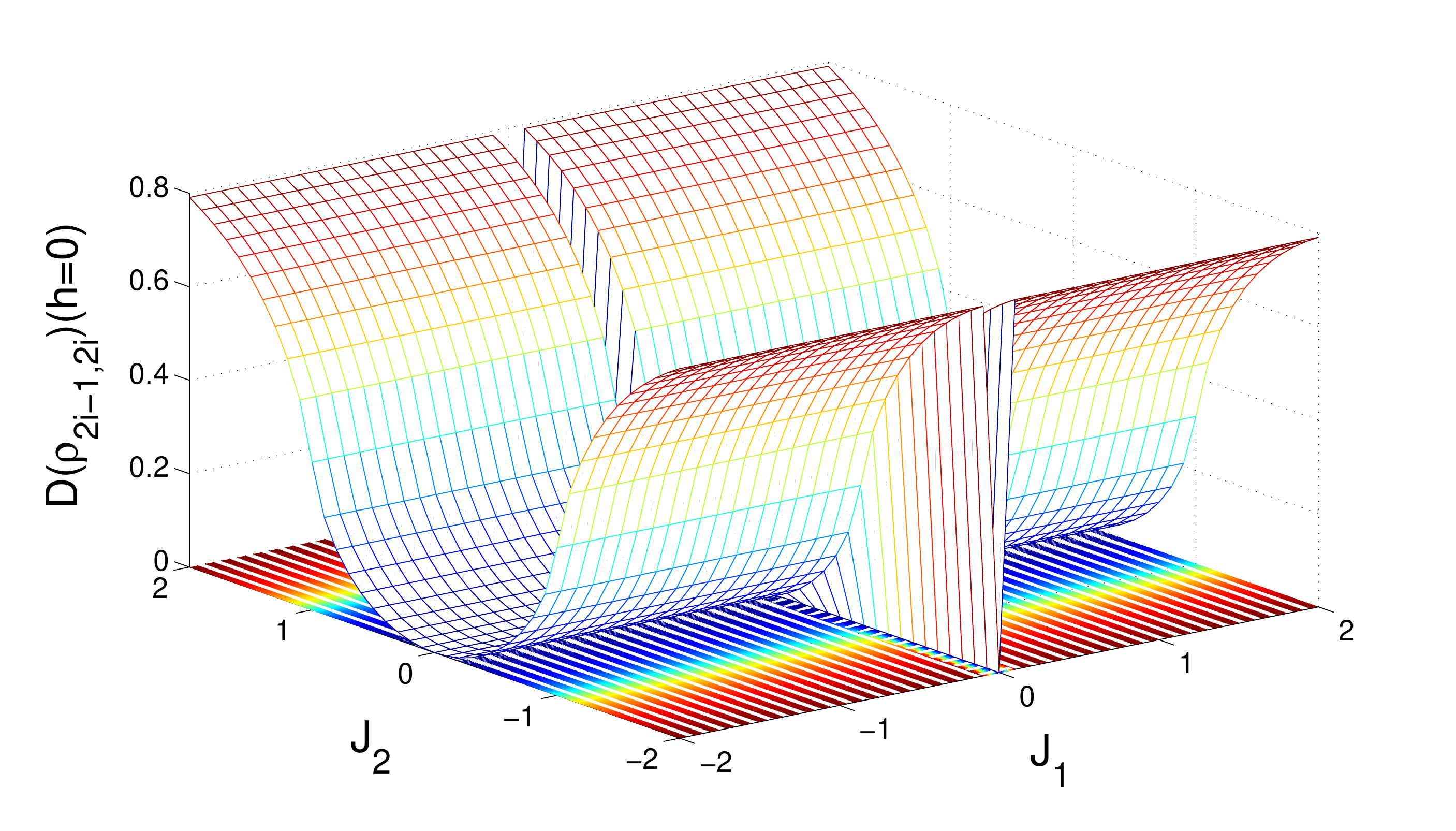}
\includegraphics[width=8cm]{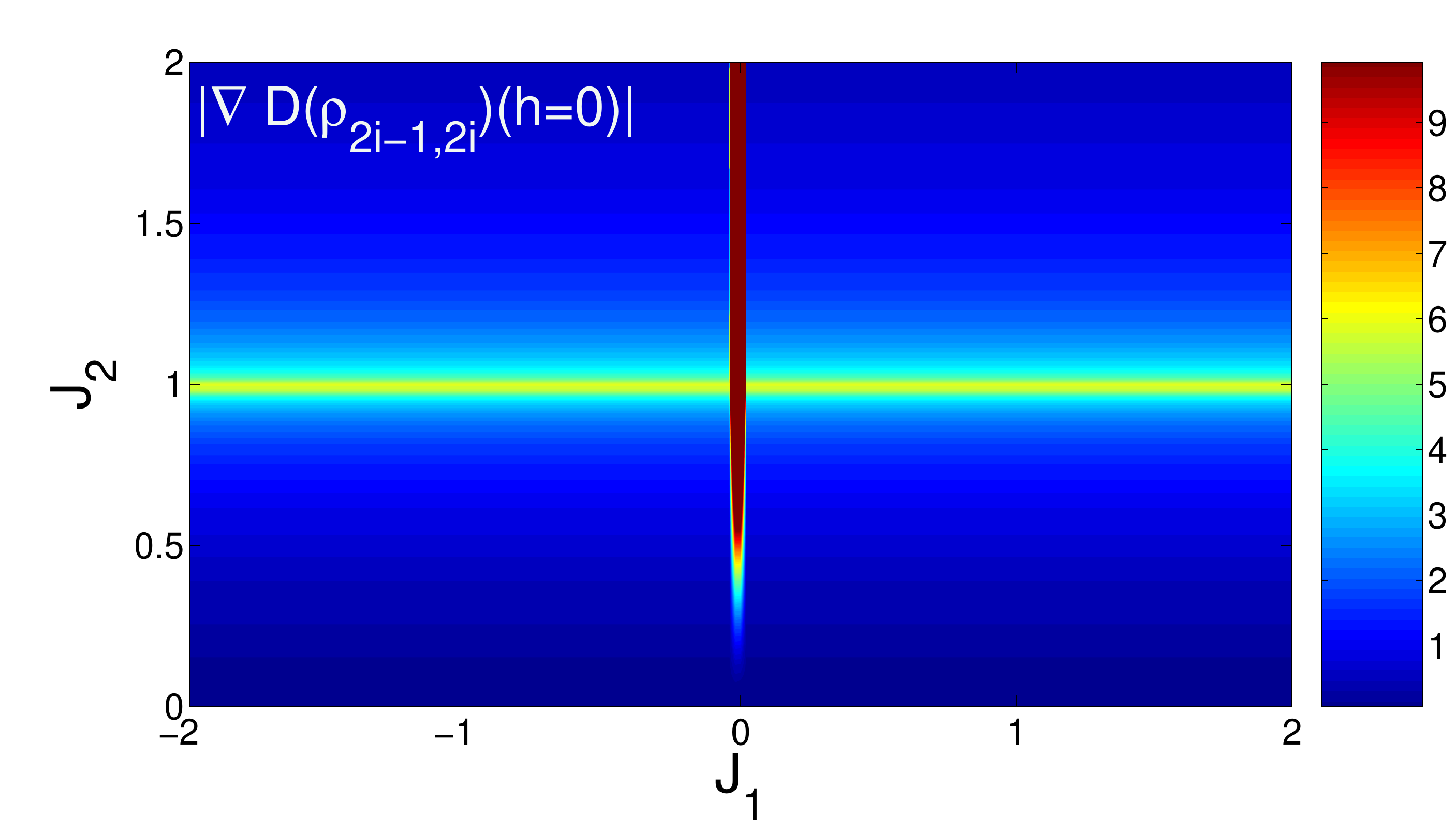}
\end{center}
\caption{(Color online) Top: QD of nearest-neighbor sites on odd
bond as functions of $J_1$ and $J_2$ with parameters $L_1 =1,L_2=0,
h=0, N=1024$. Bottom: the derivative of the QD in the $J_1-J_2$
plane. }\label{1Dcompass-J1J2h=0}
\end{figure}

\begin{figure}[t]
\begin{center}
\includegraphics[width=8cm]{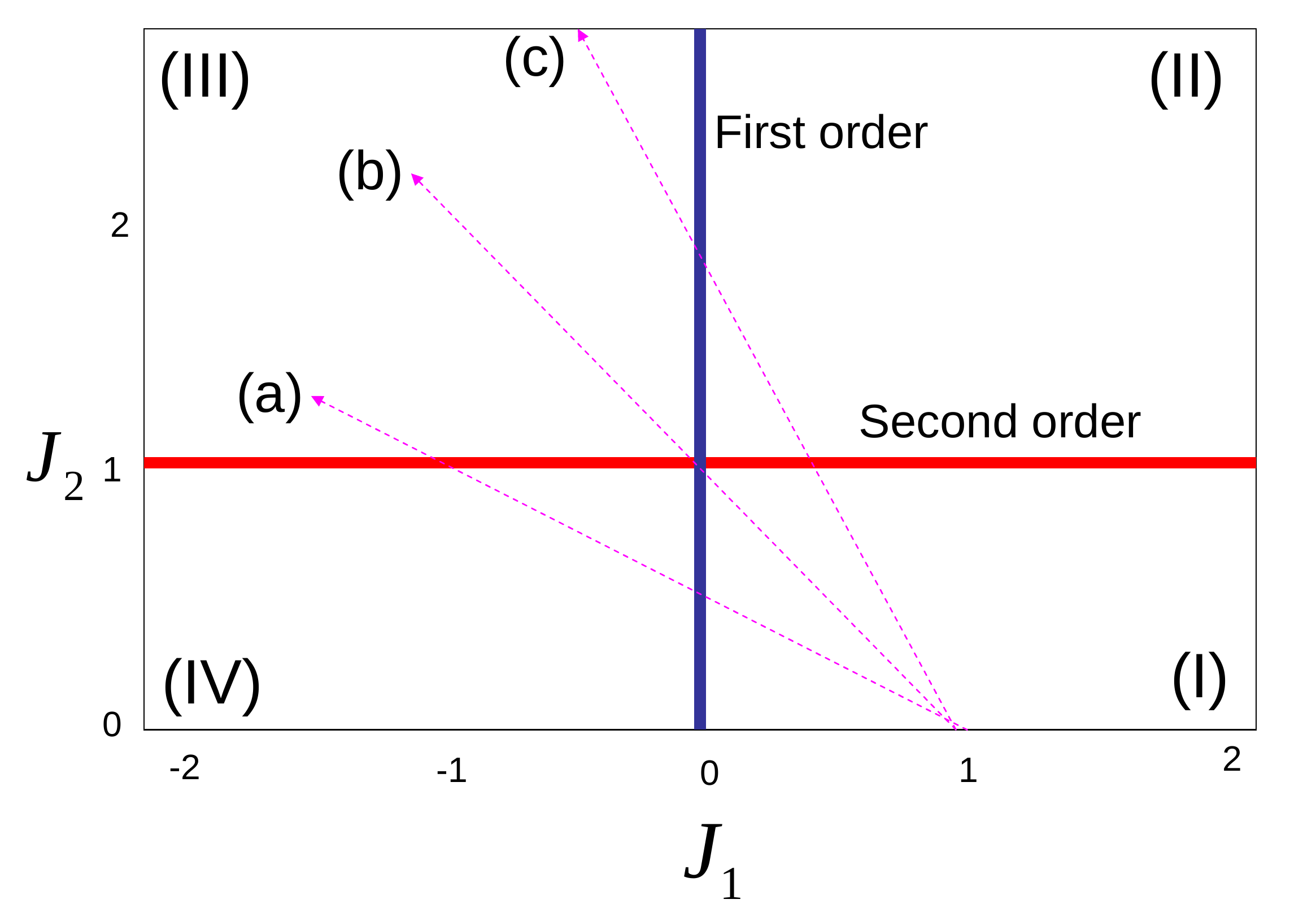}
\end{center}
\caption{(Color online) Phase diagram of the 1D quantum compass
model in the absence of magnetic field. A first-order QPT occurs at
critical line $J_1=0$ and a second-order QPT at $J_2=1$
\cite{Eriksson}. We plot QD in Fig. (\ref{threepaths}) along three
paths (a), (b), and (c) indicated by the dashed lines, which start
at the point $J_1 $=1, $J_2$=0, where the Hamiltonian
(\ref{Hamiltonian3}) reduces to that of the one-dimensional quantum
Ising model without external magnetic field.} \label{pathsofcompass}
\end{figure}

In the following, we will focus on the case of $J_2 >0$, $L_1=1$. Minus sign of $J_2$ is
irrelevant because of unitary equivalence. In absence of external
magnetic field, i.e., $h=0$, there is a first-order QPT at $J_1=0$ which
separates the phases with $\langle \sigma_{2i-1}^x \sigma_{2i}^x
\rangle=1$ from those with $\langle \sigma_{2i-1}^x \sigma_{2i}^x
\rangle =-1$. There is also a second-order QPT at $J_2=1$ separating
the phases with $\langle \sigma_{i}^y \rangle =0 $ from those with
$\langle \sigma_{i}^y \rangle \neq 0 $ \cite{Eriksson}. Two critical
lines where the energy gaps vanish, separate four gapped phases in
the parameter space, as is shown in Fig. \ref{1Dcompass-J1J2h=0}. Using Eqs. (\ref{qminfo}) - (\ref{qcorr}), the correlations can
be evaluated in detail. We observe that QD indeed
vanishes for $J_1=0$. The carmine contour line (very large values beyond the scope of colorbar) implies a first-order critical line while yellow
contour line implies a second-order critical line. The occurrence of discontinuous dips at
$J_1=0$ implies a first-order critical line and a singularity of
the first-order derivative of the QD reveals a second-order phase
transition. To interpret the relation between the QD and QPTs more clearly,
let us examine QPTs along three paths which start at the point $J_1
$=1, $J_2$=0, where Hamiltonian reduces to that of the
1D quantum Ising model. See Fig. \ref{pathsofcompass} for an illustration.

\begin{figure}[t]
\begin{center}
\includegraphics[width=8cm]{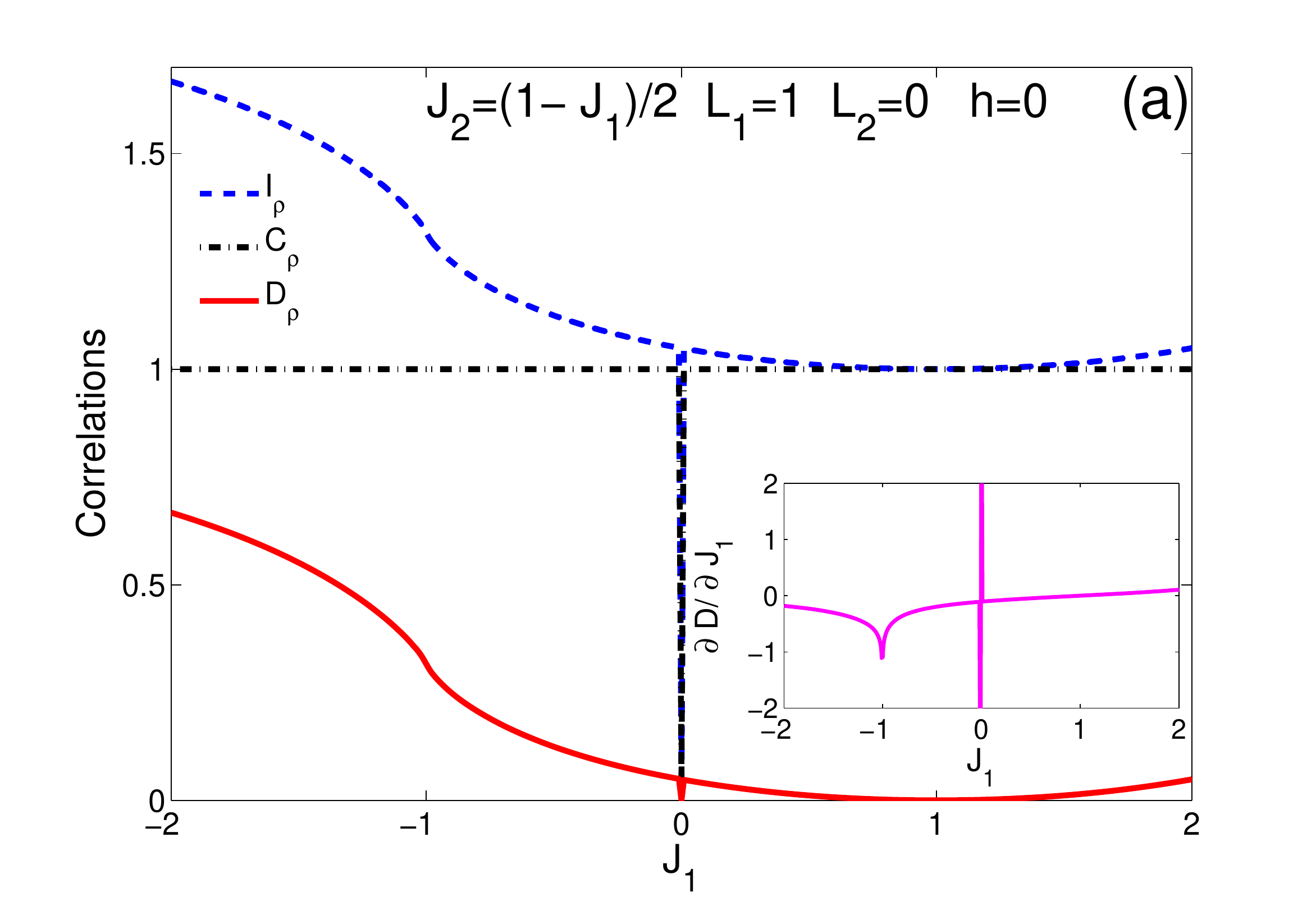}
\includegraphics[width=8cm]{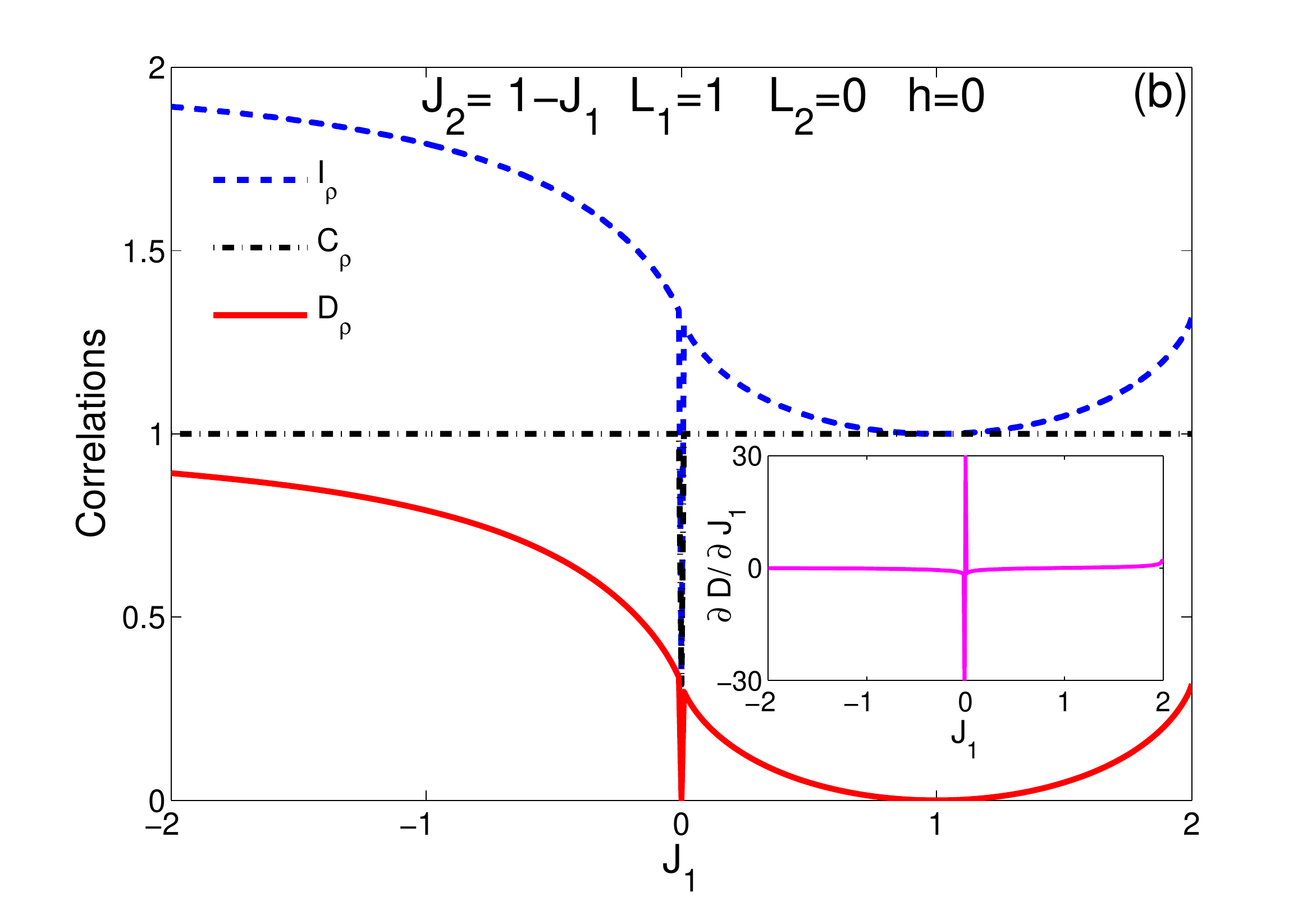}
\includegraphics[width=8cm]{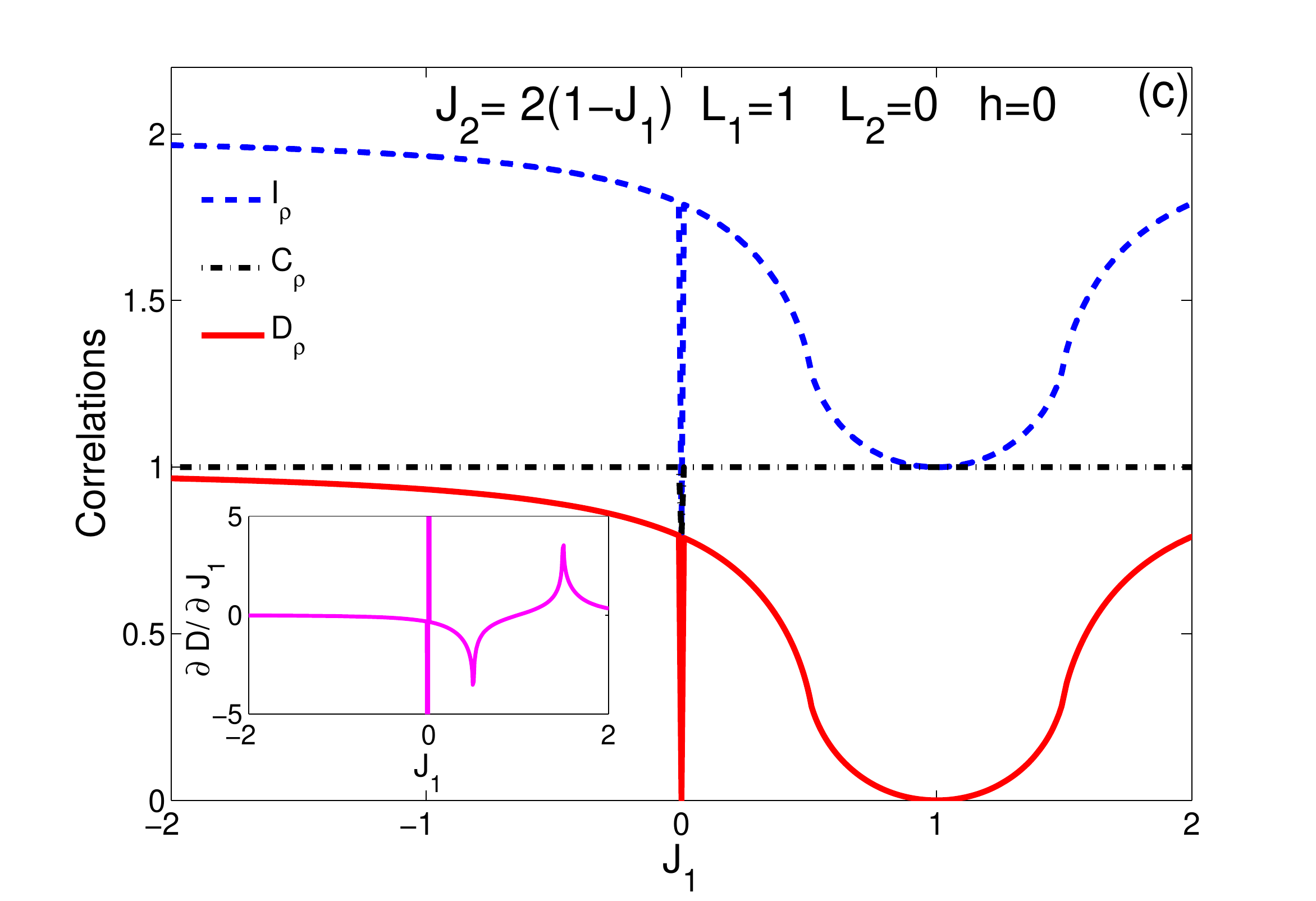}
\end{center}
\caption{(Color online) (a) The total mutual correlation, the
classical correlation and quantum discord along the paths (a)
$J_2=(1-J_1)/2$  (b) $J_2=1-J_1$ (c) $J_2=2(1-J_1)$ shown in Fig.
(\ref{pathsofcompass}) with parameters $L_1 =1,L_2 =0, N=1024$. The
insets show the corresponding first derivative with respect to
$J_1$. The diverging discontinuities have exceeded the $y$-limits of
the coordinate axes. } \label{threepaths}
\end{figure}

In Fig. \ref{threepaths}, we plot the QD verse $J_1$ along the three paths shown in Fig. \ref{pathsofcompass}. It is clear
that QD exhibits a sudden drop at $J_1=0$, suggesting a first-order
QPT. For paths $J_2=(1-J_1)/2$ and
$J_2=2(1-J_1)$, a diverging first derivative of QD signals a second-order QPT.
 As for the multicritical point on the line
$J_2=1-J_1$, a first-order phase transition dominates indicated by a
discontinuity of the QD. Note that the classial correlation $C(\rho_{ij})=1$ at $J_1 \neq 0$, and also has
an abrupt decrease at $J_1=0$.

\begin{figure}[t]
\begin{center}
\includegraphics[width=8cm]{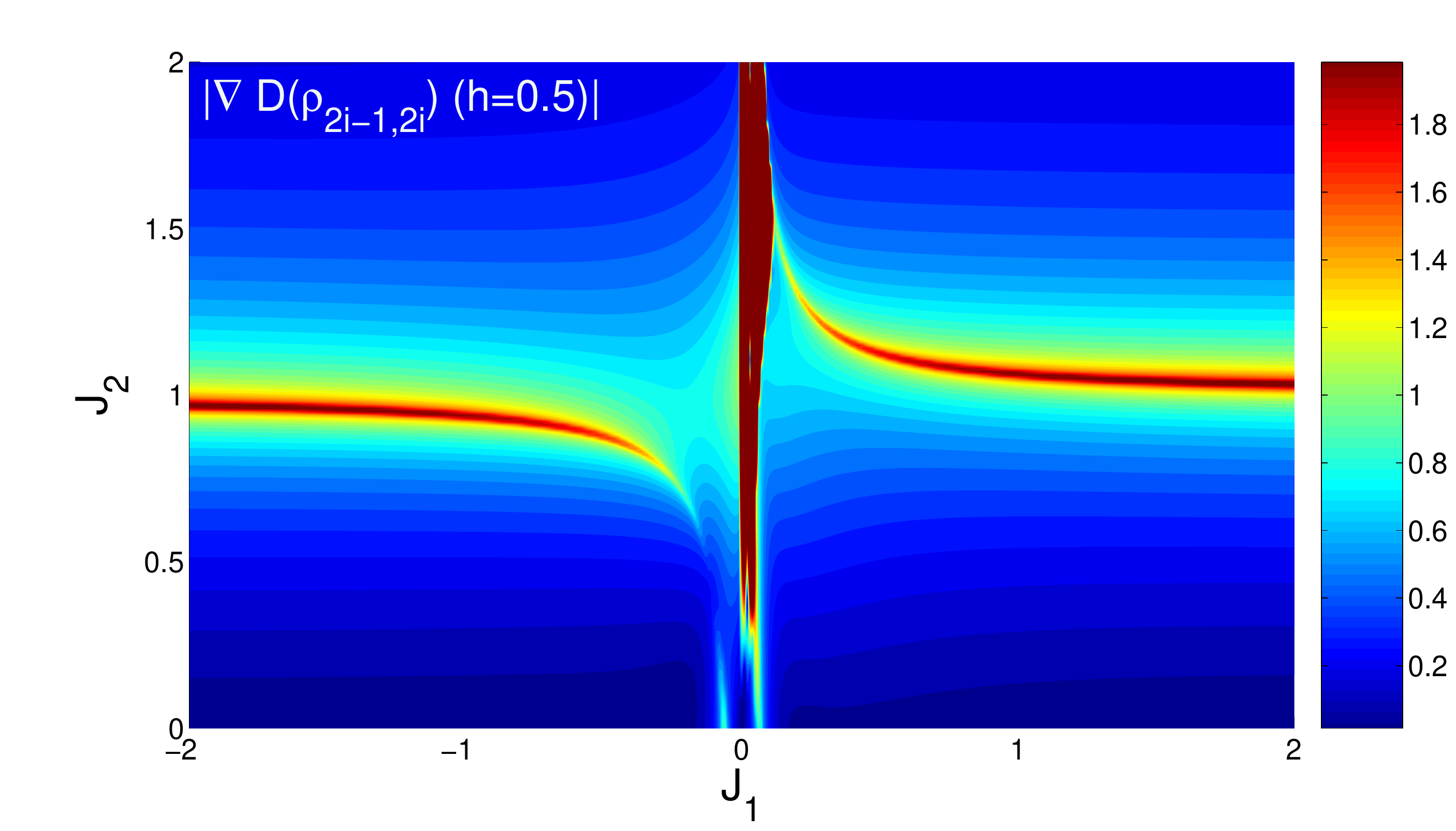}
\includegraphics[width=8cm]{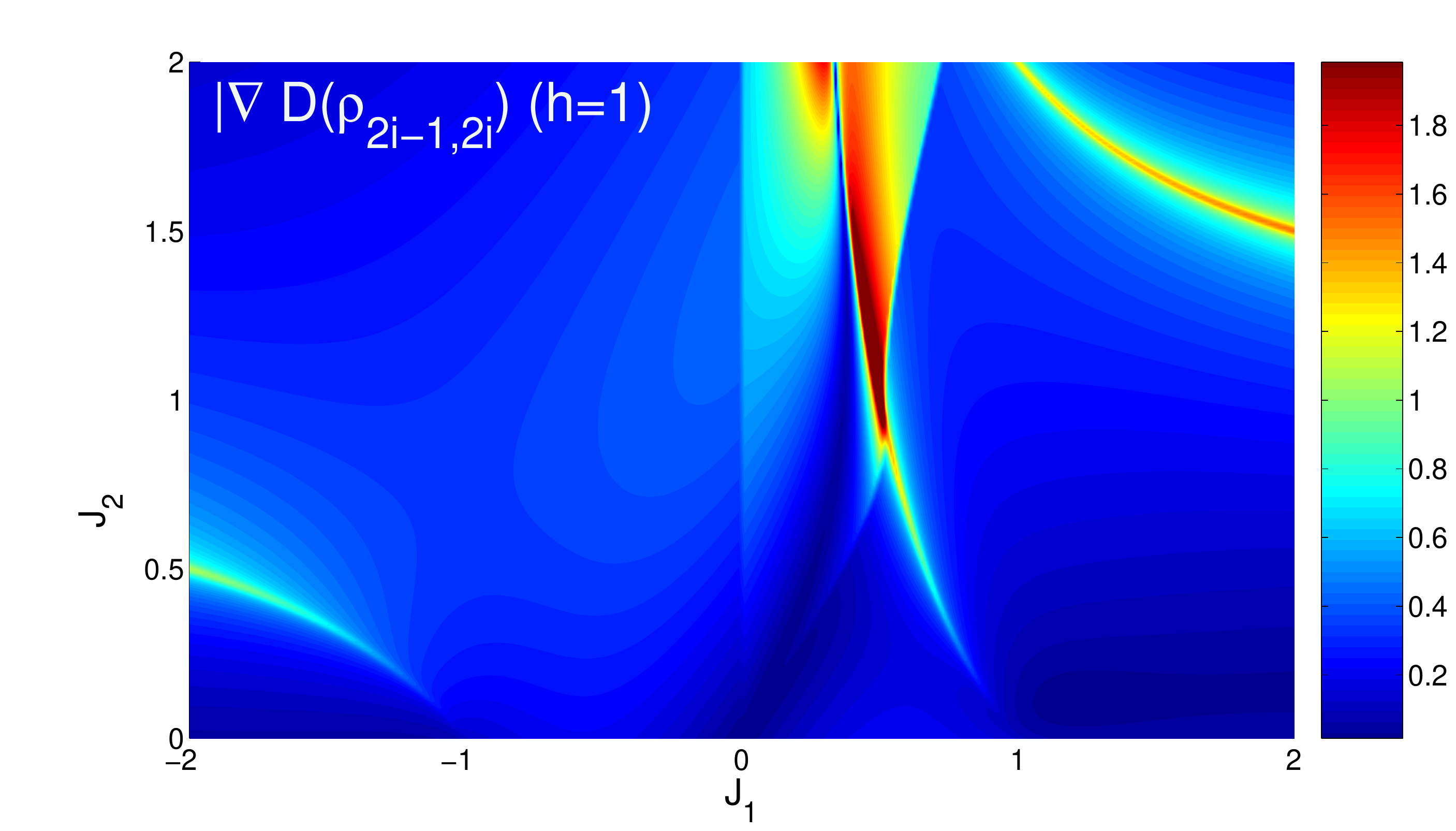}
\end{center}
\caption{(Color online) The derivative of the QD in the $J_1-J_2$
plane when $L_1 =1,L_2 =0, N=1024$ for magnetic field $h=0.5$ and
$h=1$ respectively. }\label{1Dcompass-J1J2}
\end{figure}

As mentioned above, the zero discord state is incredibly fragile, and finite
perturbation can create QD. For example, the magnetic field will lift the high degeneracy of the 1D
compass model. Therefore, we are especially interested
in the role of the QD in detecting QPTs in 1D
extended compass model under the magnetic field. In Fig. \ref{1Dcompass-J1J2}, we plot the QD of nearest-neighbor qubits
as the function of $J_1$ and $J_2$ with external magnetic field. We
present the contour map of the first derivative of the QD. With the increase
of magnetic field $h$, we find that the critical lines deviate from those for $h=0$,
and the second-order critical line develops into hyperbolas.

\begin{figure}[t]
\begin{center}
\includegraphics[width=8cm]{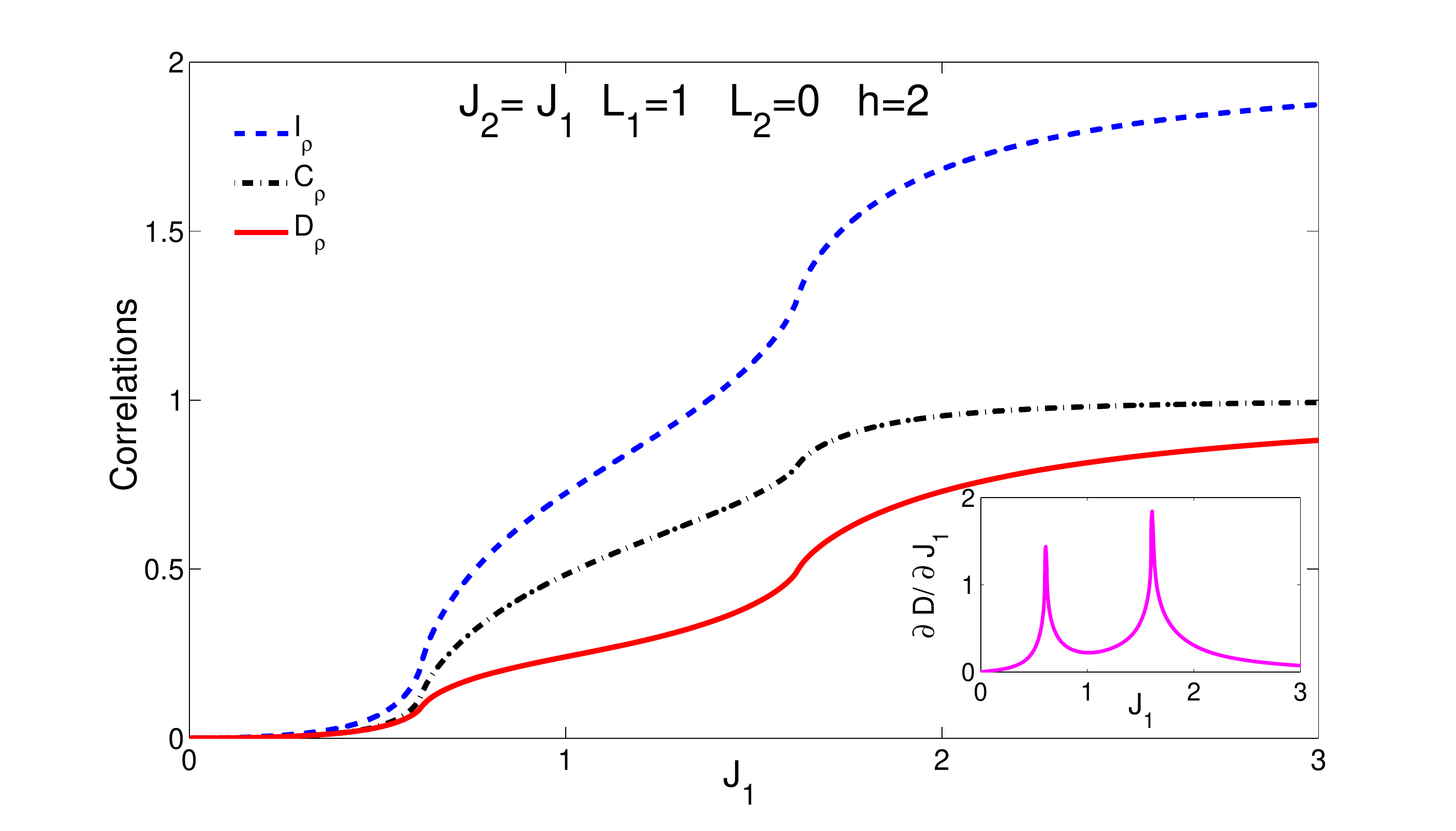}%
\end{center}
\caption{(Color online) The total mutual correlation, the classical
correlation and quantum discord along the path $J_2=J_1$ when $L_1
=1,L_2 =0, L=2048$ for magnetic field $h=2$. The inset shows its
first derivative.} \label{J2=J1}
\end{figure}

\begin{figure}[t]
\begin{center}
\includegraphics[width=8cm]{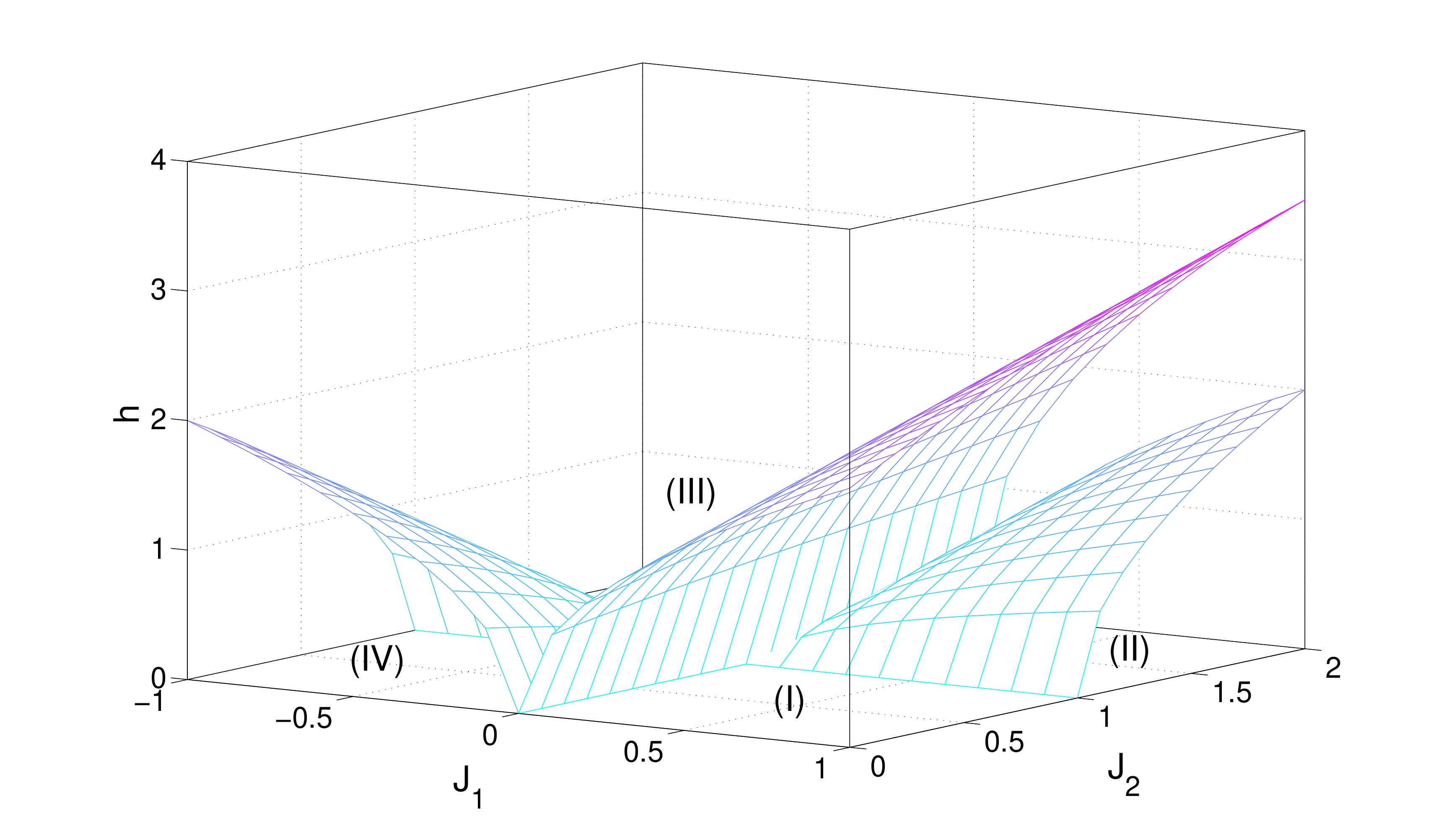} %
\end{center}
\caption{(Color online) The phase diagram of the extended compass
model in the transverse magnetic field, $L_1 =1,L_2 =0$. (I)
spin-flop phase for $J_1 >$ 0, $J_2 <$ 1; (II) antiparallel ordered
of spin $y$ component for $J_1 >$ 0, $J_2 >$ 1; (III) saturated
ferromagnetic phase for $J_1 <$ 0, $J_2 >$ 1; (IV) stripe
antiferromagnetic phase for $J_1 <$ 0, $J_2 <$ 1.
}\label{Phasediagramof1Dcompass}
\end{figure}

Furthermore, we plot QD as a function of $J_1$ along the path $J_2=J_1$ and its corresponding
first-derivative for $h =2$ in Fig. \ref{J2=J1}. The divergent
peaks show clearly that there will be QPTs at $J_{1c}=(L_1 \pm
\sqrt{ L_1 ^2 +h^2} )/2 $ and $(-L_1 \pm \sqrt{ L_1 ^2 +h^2})/2$, and these QPTs belong to second order.
Hence, the phase diagram of 1D EQCM in the transverse magnetic field is sketched in Fig. \ref{Phasediagramof1Dcompass},
which is identical with that analyzed by correlation functions
\cite{arXiv1101.3673}. It is worthy noting that there are only
continuous phase transitions once applying the magnetic field.

\section{Entanglement}
\label{Entanglement} Given the exact solution of the 1D EQCM, we
have a rare opportunity to analytically probe the entanglement
in the ground state of a complex system. We
here focus on one of the most frequently used entanglement measure:
concurrence \cite{Wootters}. The concurrence can quantify
entanglement for any bipartite system that relates to the two-site
RMD $\rho_{ij}$, which is defined as
\begin{eqnarray}
C(\rho_{i j})=\max\{ 0,  \lambda_1  -  \lambda_2
- \lambda_3  - \lambda_4 \},
\end{eqnarray}
where $\lambda_i$ are the eigenvalues in decreasing order of the
auxiliary matrix
\begin{eqnarray}
\zeta=\sqrt{\rho_{{i}{j}}(\sigma_{i}^y
\sigma_{j}^y)\rho_{{i}{j}}^*(\sigma_{i}^y \sigma_{j}^y)}.
\end{eqnarray}
Here $\rho_{{i}{j}}^*$ denotes the complex conjugation of
$\rho_{{i}{j}}$ in the standard basis. The eigenvalues of $\zeta$ are
\begin{eqnarray}
\lambda_{1,2}&=&\frac{1}{4} \Big\vert \sqrt{1+\langle \sigma_i^z \rangle +
\langle \sigma_j^z \rangle + \langle \sigma_i^z \sigma_j^z  \rangle}
\nonumber \\
&\times&\sqrt{1-\langle \sigma_i^z \rangle - \langle \sigma_j^z
\rangle + \langle \sigma_i^z \sigma_j^z  \rangle} \pm \vert \langle
\sigma_i^x \sigma_j^x  \rangle - \langle \sigma_i^y \sigma_j^y
\rangle \vert \Big\vert, \nonumber\\
\lambda_{3,4}&=&\frac{1}{4} \Big\vert \sqrt{1+\langle \sigma_i^z \rangle-
\langle \sigma_j^z \rangle - \langle \sigma_i^z \sigma_j^z  \rangle}
\nonumber \\
&\times&\sqrt{1-\langle \sigma_i^z \rangle + \langle \sigma_j^z
\rangle - \langle \sigma_i^z \sigma_j^z  \rangle} \pm \vert \langle
\sigma_i^x \sigma_j^x  \rangle + \langle \sigma_i^y \sigma_j^y
\rangle \vert \Big\vert.\nonumber \\
\end{eqnarray}

\begin{figure}[t]
\begin{center}
\includegraphics[width=8cm]{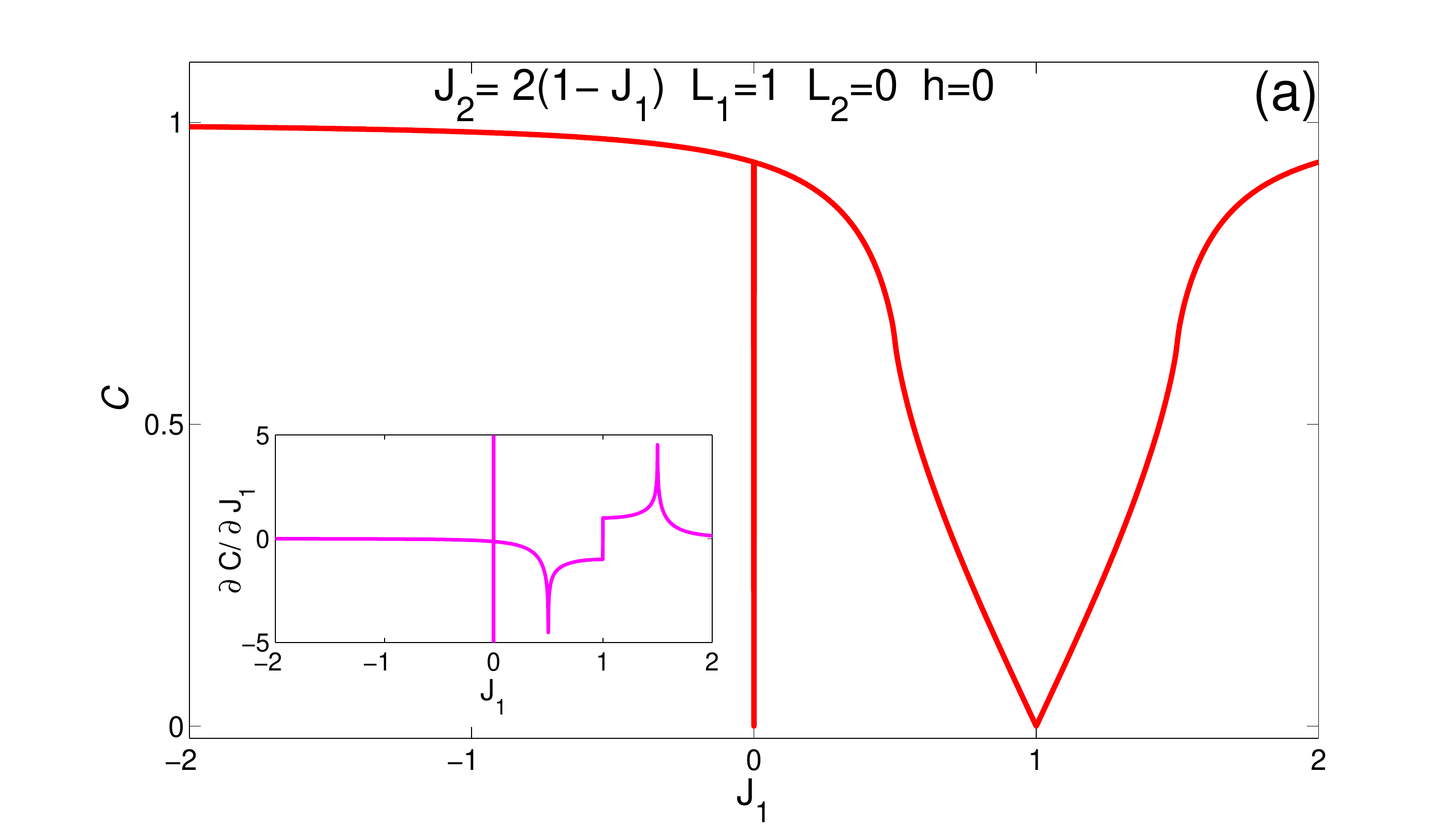}
\includegraphics[width=8cm]{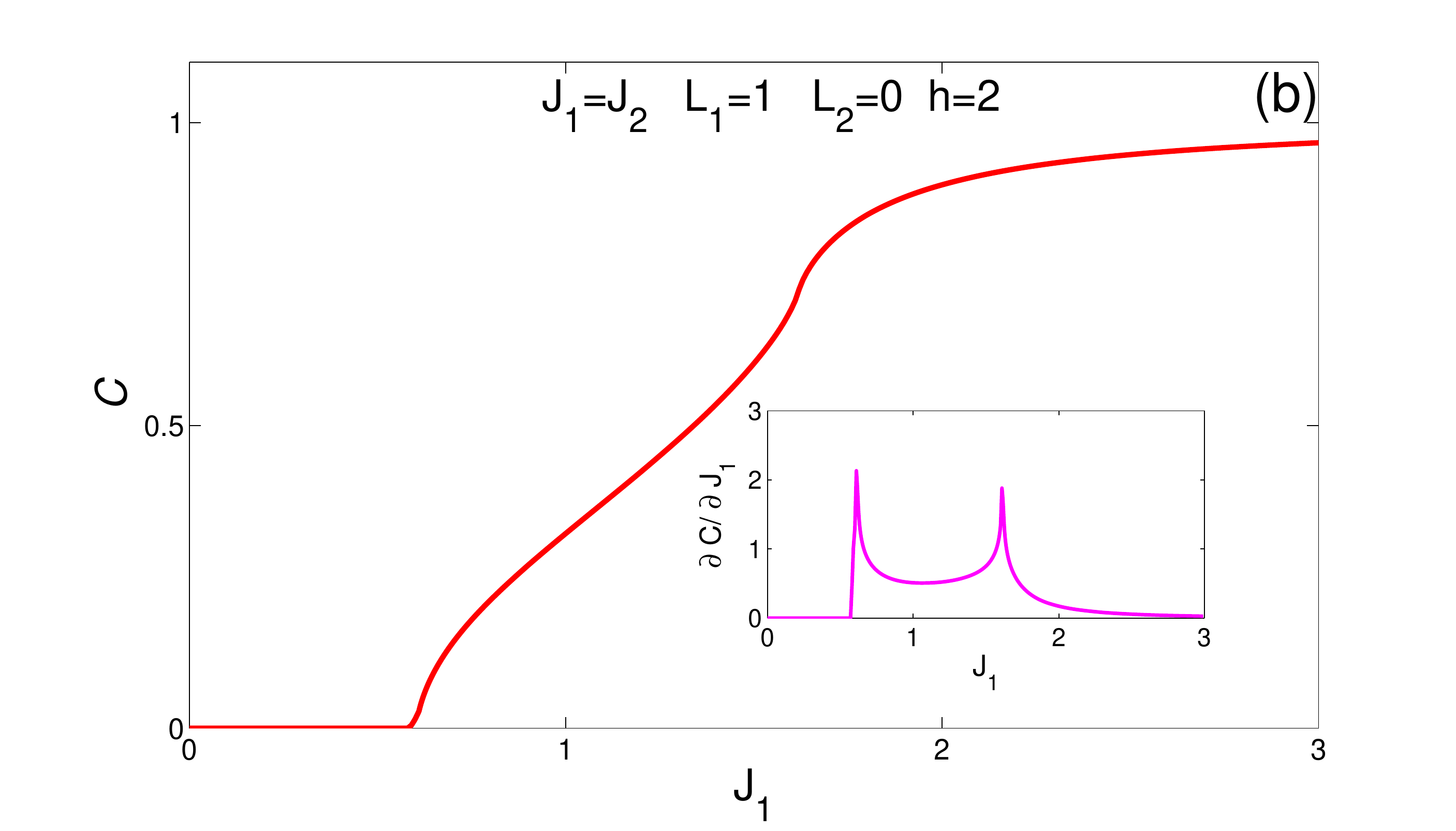}
\end{center}
\caption{(Color online) The concurrence of nearest neighbor spins on
the odd bonds of with respect to $J_1$ along path (a)$J_2=2 (1-J_1)$
(b) $J_2=J_1$ with $L_1 =1,L_2 =0, N=1024$. The corresponding first
derivatives of concurrence are displayed in insets.}
\label{Concurrence-J1=J2}
\end{figure}

For two spins that are on the same odd bond $\{2i-1,2i\}$, when
$J_1>0$, $L_1=1$, $L_2=0$, the ground state resides in the subspace,
where $\langle\sigma_i^z\rangle=0$, $\langle\sigma_{2i-1}^x
\sigma_{2i}^x\rangle=-1$, $\langle\sigma_{2i-1}^y
\sigma_{2i}^y\rangle=\langle\sigma_{2i-1}^z \sigma_{2i}^z\rangle$,
and then it is immediately clear that
$C(\rho_{ij})=-\langle\sigma_{2i-1}^y \sigma_{2i}^y\rangle$. On the
other hand, when $J_1<0$, the ground state is in another subspace, where
$\langle\sigma_{2i-1}^y \sigma_{2i}^y\rangle=-\langle\sigma_{2i-1}^z
\sigma_{2i}^z\rangle$, $\langle\sigma_{2i-1}^x
\sigma_{2i}^x\rangle=1$, and then
$C(\rho_{ij})=-\langle\sigma_{2i-1}^y \sigma_{2i}^y\rangle$.
However, at $J_1=0$, $\langle\sigma_i^z\rangle=0$,
$\langle\sigma_{2i-1}^x \sigma_{2i}^x\rangle=0$,
$\langle\sigma_{2i-1}^x \sigma_{2i}^x\rangle=0$,
$\langle\sigma_{2i-1}^z \sigma_{2i}^z\rangle=0$, and then
$C(\rho_{ij})=0$, which does not support the conjecture in Ref.
\cite{Eriksson} that the first-order QPT at $J_1$ =0 is not signaled
by the pairwise concurrence. As is displayed in Fig.
 \ref{Concurrence-J1=J2}, the concurrence of nearest-neighbor
pairs of spins captures the discontinuity across the first-order
transition point when there is no external magnetic field, and
the diverging peaks of the first derivative of concurrence imply the
continuous QPTs.

\section{Discussion and conclusion}
\label{Conclusion}

Despite the resemblance in characterizing the
QPTs in 1D EQCM, there still exist much difference between the
concurrence and the QD. For example, the QD can
present the correlations between neighbors farther than the
next-nearest, while pairwise entanglement may be absent for these neighbors
\cite{PhysRevA.82.012106}. Besides, the QD can characterize QPTs by
exhibiting long-range decay as a function of distance in spin
systems, which is different from the behavior of pairwise
entanglement \cite{arXiv1012.5926}. Moreover, the thermal
fluctuations extinguish the entanglement, while the QD is robust to
spotlight QCPs at finite temperature \cite{Mazzola}. The QD even
increases with temperature in some cases \cite{Werlang2}.
Furthermore, there is an evidence that QD may present a scaling law,
which is not the case for entanglement \cite{Tomasello}. General
speaking, quantum correlations are more fundamental than quantum
entanglement, and may reveal more information about the quantum systems.

In conclusion, we have examined pairwise QD by exactly solving the 1D EQCM in the presence of an external transverse
magnetic field. We find that the QD is equal to zero for the 1D
compass model. Remarkably, we have successfully extracted information of the location and the
order of the QPTs in 1D EQCM by consideration of the derivative of
the QD with respect to the coupling parameters. We conclude that a
first-order QPT is associated with a discontinuity in the QD and a
continuous second-order transition demonstrates a diverging first
derivative of QD. The mixed first-order and second-order phase
transition point features a discontinuity. We get the analytic
expressions of critical magnetic fields for the field-induced QPTs, which
are of second order. As a result, we then obtain the phase diagram of 1D EQCM in the transverse
magnetic field. For comparisons, we show that
the pairwise concurrence can characterize the phase transitions by
exhibiting similar behaviors. Nevertheless, the QD is believed to be
more fundament than concurrence in quantifying the quantum correlations, and should be of general interest in future studies.

\section{Acknowledgements}

Wen-Long You acknowledges the support of the Natural Science
Foundation of Jiangsu Province under Grant No. 10KJB140010 and the
National Natural Science Foundation of China under Grant
No.11004144.

\end{document}